\documentclass{vgtc}                          

\ifpdf
  \pdfoutput=1\relax                   
  \usepackage{graphicx}                
  \DeclareGraphicsExtensions{.pdf,.png,.jpg,.jpeg} 
\else
  \usepackage{graphicx}                
  \DeclareGraphicsExtensions{.eps}     
\fi%

\graphicspath{{figures/}{pictures/}{images/}{./}} 

\usepackage{microtype}                 
\PassOptionsToPackage{warn}{textcomp}  
\usepackage{textcomp}                  
\usepackage{mathptmx}                  
\usepackage{times}                     
\usepackage{cite}                      
\usepackage{tabu}                      
\usepackage{booktabs}                  

\onlineid{0}

\vgtccategory{Research}

\vgtcinsertpkg

\title{How do Visualization Designers Think? Design Cognition as a Core Aspect of Visualization Psychology}

\author{Paul Parsons\thanks{e-mail: parsonsp@purdue.edu}\\ %
        \scriptsize Purdue University }

\abstract{There are numerous opportunities for engaging in research at the intersection of psychology and visualization. While most opportunities taken up by the VIS community will likely focus on the psychology of users, there are also opportunities for studying the psychology of designers. In this position paper, I argue the importance of studying \textit{design cognition} as a necessary component of a holistic program of research on visualization psychology. I provide a brief overview of research on design cognition in other disciplines, and discuss opportunities for VIS to build an analogous research program. Doing so can lead to a stronger integration of research and design practice, can provide a better understanding of how to educate and train future designers, and will likely surface both challenges and opportunities for future research. %
} 

\begin{document}


\firstsection{Introduction}

\maketitle
Psychology is undoubtedly important for visualization research and practice. Numerous scholars have pointed to the need for better integration of the two fields, especially as an expansion beyond perception to focus on higher-level cognitive processes and structures. However, virtually all calls for more of this kind of research focus \textit{entirely on users and not on designers}. Although more research on user psychology is indeed necessary, it can provide only a partial view of the ways in which visualizations are created and used. Most disciplines with a robust relationship between research and practice have developed research programs investigating how designers think while engaged in their design practice, commonly referred to as \textit{design cognition}. These include both longstanding fields like architecture \cite{lawson_cognitive_1979} and emergent fields like instructional design \cite{perez_designer_1995}, graphic design \cite{stones_seeing_2010}, and user experience design \cite{Gray_2016}. These disciplines have all recognized the value of studying how designers think and know as a valid and useful complement to studying the psychological aspects of the use of artifacts. The integration of these two perspectives can lead to a more holistic view of how visualizations are created and used.

Design research in VIS has largely been model driven, where attempts are made to codify aspects of design in abstract forms. Popular examples include process models (e.g., Design Study Methodology \cite{sedlmair_design_2012}, Design Activity Framework \cite{mckenna_design_2014}) and decision models (e.g., Nested Blocks \cite{munzner_nested_2009}, Nested Blocks and Guidelines \cite{meyer_nested_2015}). While the latter models focus on decisions that designers can or should make in design situations, they do not engage with decision making at the level of cognitive processes or structures. Thus there is a gap in our understanding of how and why design decisions are made, and the psychological factors that influence their outcomes.

In this position paper, I argue that there is a need to focus on \textit{design cognition} in VIS research. I provide a brief overview of research on design cognition in other disciplines, and discuss opportunities for VIS to build an analogous research program. Doing so can lead to a stronger integration of research and design practice, can provide a better understanding of how to educate and train future designers, and will likely surface both challenges and opportunities for future research.

\section{Design Cognition}
How do designers formulate and solve design problems? What kinds of cognitive processes do they rely on while doing so? These are the types of questions asked by researchers studying design cognition. Rather than foregrounding the methods, tools, or outcomes of designers, studies in design cognition investigate how and why designers think the ways they do while designing. Design cognition has been studied across a wide variety of domains, including engineering\cite{atman_engineering_2007}, architecture \cite{lawson_cognitive_1979}, computer science \cite{carroll_scenarios_2002}, instructional design \cite{perez_designer_1995}, and graphic design \cite{stones_seeing_2010}. Across these disciplines, many aspects of cognition in design have been investigated, including, among others, episodic memory \cite{lawson_context_2001}, fixation \cite{purcell_design_1996}, chunking \cite{mao_evidence_2020}, bias \cite{damle_biasing_2009}, abductive reasoning \cite{cramer-petersen_empirically_2019}, analogical reasoning \cite{visser_two_1996}, metacognitive monitoring and control \cite{ball_advancing_2019}, and recall \cite{crismond_investigate-and-redesign_1997}. 

Much of the research on design cognition is concerned with how designers navigate the complexity and uncertainty of real-world design situations \cite{buchanan_wicked_1992,Stolterman2008,Simon_1969}. A number of core strategies have been identified through empirical investigation, including conjecture-based problem formulation, problem-solution co-evolution, analogical reasoning, mental simulation, and ﬁxated solution generation \cite{ball_advancing_2019,Eastman2001}. Many of the cognitive processes that are relevant for studying the use of visualizations are also important for understanding their design. For instance, studies have shown that designers rely on chunking to ideate effectively \cite{mao_evidence_2020}, employ abductive reasoning during concept selection \cite{dong_effect_2015}, are influenced by color in ways that bias their thinking while sketching \cite{damle_biasing_2009}, and struggle with fixation while generating ideas \cite{Crilly_2019}. Given the considerable evidence of such issues impacting design across numerous disciplines, it is very likely that these are also important for understanding how and why visualization designers design the ways they do.

Although it may be tempting to simply borrow these findings from these other disciplines and apply them to VIS, research has shown that significant differences exist in design cognition across domains, even though there are similarities \cite{visser_design:_2009}. For instance, Akin \cite{akin_variants_2001} notes significant differences in design cognition among engineers and architects, and Purcell and Gero \cite{purcell_design_1996} identifies differences between mechanical engineers and product designers. In a review paper examining design across numerous domains, Visser \cite{visser_design:_2009} affirms both these differences and similarities, and speculates that these differences may have implications for the kinds of knowledge that designers rely on. 

Following this work, it is reasonable to assume that design cognition in VIS will share similarities with these other fields, yet will also have its own unique characteristics. For instance, designers in other fields do not deal with issues regarding data, mapping abstract information onto visual forms, and interactivity in the ways that VIS designers must. Research on design cognition has significantly influenced theory, practice, and education in numerous design fields \cite{jiang_protocol_2009}, and could similarly do so for VIS. However, the particular facets of VIS that make it different from engineering or graphic design, for instance, must be carefully examined as a part of such an effort.



\subsection{Design and Applied Research}
One possible reason why VIS has not seen a focus on design cognition, and design practice more broadly, is a common assumption that design is simply an application of knowledge generated from scientific research. This view suggests that if designers know enough---if they understand the principles and concepts that come from research---then they can apply them in their work. For instance, if enough studies are done on visual encoding and perception, or memorability and embellishment, or bias and chart types, designers just have to know the results and determine how to apply them in context. 

This view was prominent in multiple design fields decades ago, but has since largely been abandoned by design scholars, as it does not accurately reflect the true nature of design practice \cite{cross_designerly_1982,lawson_how_2006}. While scientific knowledge certainly plays a role in design, it is not sufficient for good design \cite{Stolterman2008}. Rather, designers rely on a host of personal and situated factors, along with more formal types of knowledge, to engage appropriately with the complexity of design practice \cite{Parsons_judgement_2020}. Buchanan \cite{buchanan_wicked_1992} articulates how widespread this assumption has been, noting that ``each of the sciences that have come into contact with design has tended to regard design as an `applied' version of its own knowledge'', emphasizing the mistake of viewing design as simply a ``practical demonstration'' of scientific findings. Thus, even if a robust program of research at the intersection of psychology and visualization is developed, if its scope is limited to users only---and especially if design is viewed as an application of research findings---we will fail to understand how to influence design practice effectively.

\section{Opportunities and Research Questions}
Because the study of design cognition has a rich history in other disciplines, yet is still nascent in VIS, research questions can be translated from fields like architecture, instructional design, and others, and posed in a VIS context. For instance, much research has focused on the differences between novice and expert designers, and research questions can be generated to investigate these differences, including: How do novices and experts differ with respect to framing problems involving complex datasets and use cases? Are experts capable of more flexible ways of recognizing and framing problems? How do novices and experts differ with respect to chunking while generating ideas for chart types? Do experts engage in more advanced chunking strategies that allow for nuanced application of principles regarding visual encodings or embellishments? 

Another line of research can select individual concepts or topics to study. For instance, with respect to creativity, the ways in which cognitive strategies are used during ideation, and how designers may get fixated on certain things, can be examined. Given an example of a well-known visualization for a particular kind of data and context, do designers become fixated on that particular type of solution, unable to see viable alternatives? And what kinds of supports can be given to mitigate that fixation? 


Another line of research may characterize the similarities and differences among VIS and other fields with respect to design cognition. For instance, do VIS designers rely on abductive reasoning in similar ways to instructional  designers? Is mental simulation different for VIS designers and UX designers? And what kinds of metacognitive strategies do VIS designers rely on that may be similar or different to software engineers?

A different approach can start with the more universal processes of design cognition, using those to investigate differences in novices and experts or specific cognitive processes and structures. In his seminal work, Cross \cite{Cross2001,Cross2001a} articulated two fundamental processes of design cognition: problem formulation and solution generation. These two processes are fairly abstract, yet are essential aspects of design and are thus mostly universal. Problem formulation refers to processes in which designers make sense of ill-defined situations and determine the `problem' and its implications. Solution generation refers to processes in which designers move from a problem to a satisfactory solution. This could be a reasonable starting point for design cognition research in VIS. Designers could be recruited and given a problem brief comprising a dataset, target users, and set of tasks that users need to accomplish. Their verbalizations could be captured and analyzed, with specific attention paid to known indicators of these cognitive processes. For instance, it is generally recognized that during problem formulation designers attend to specific aspects of the problem space by naming them, articulating relevant concerns as part of the problem frame, and articulating a coherent narrative that helps guide subsequent design decisions \cite{schon_reflective_1983}. 

The possibilities noted here are only a small sample, and it is certain that this list is far from exhaustive. There are likely dozens of research questions that can be posed based on prior studies on other areas. However, the ideas mentioned here are simply a starting point to begin a discussion on engaging in this kind of research.


\subsection{Research Approaches}
Previous research has heavily relied on ``protocol studies'' to investigate the nature of design cognition \cite{Cross2001,Eastman2001}. This method, which is already well-known to VIS researchers doing human-subjects studies, involves asking designers to `think-aloud' while doing a design activity. These studies generate verbal protocols that can be transcribed and analyzed with the goal of uncovering aspects of thinking and reasoning. This kind of approach can be taken with individual designers who work alone on design problems, or with teams of designers working together. Team-based protocols can be used to elicit socio-cognitive facets of collaborative design cognition.

Similar to other aspects of VIS research, designers can be studied in both controlled settings, such as a lab or workshop, and in less controlled settings, such in their everyday design settings. Studies in controlled settings are beneficial as they allow common design tasks to be given to participants, and allow for the control of variables, including time spent, access to resources, and so on. Although empirical lab studies are commonly employed in design research (e.g., \cite{cash_comparison_2013,hernandez_understanding_2010}), they differ from realistic practitioner contexts in a number of ways. For instance, lab studies may exclude factors that shape design work in commercial settings, including the effects of organizational culture, project timescales, project management and workload. Lab studies may also present participants with relatively simple problems over short time periods, which are not often representative of real-world design tasks. As is often the case in experimental research, there is the risk of reducing both ecological and external validity \cite{cash_comparison_2013}. For these reasons, it is beneficial to conduct studies in both controlled ``lab'' settings and ``in the wild'' of real-world practice. Studies can employ a range of methods, including protocol analysis, semi-structured interviews, diary studies, contextual observations, and co-design workshops.


\section{Summary}
There are numerous opportunities for engaging in research at the intersection of psychology and visualization. While most opportunities taken up by the VIS community will likely focus on the psychology of users, there are also opportunities for studying the psychology of designers. In this position paper, I have argued the importance of studying \textit{design cognition} as a necessary component of a holistic program of research on visualizaiton psychology. Perhaps the most obvious implication for this kind of research is in the education and training of future visualization designers. However, there are also implications for research and design practice. Understanding design cognition can help generate research topics and questions for the research community, and can stimulate the creation of design methods, concepts, and other types of design knowledge that can be used in practice.


\bibliographystyle{abbrv-doi}

\bibliography{template}
\end{document}